\documentclass[12pt]{iopart}
\expandafter\let\csname equation*\endcsname\relax
\expandafter\let\csname endequation*\endcsname\relax
\usepackage{amsfonts}
\usepackage{amsmath}
\usepackage{iopams}
\usepackage{amssymb}
\usepackage{mathtools}
\usepackage{cancel} 
\usepackage{bm}
\usepackage[usenames, dvipsnames]{color} 
\usepackage{dcolumn}
\usepackage{epic} 
\usepackage{epsfig}
\usepackage{feynmp}
\usepackage{graphicx}
\usepackage{grffile}
\usepackage[breaklinks,colorlinks = true,linkcolor = red,urlcolor=blue,citecolor=cyan]{hyperref}
\usepackage{mathrsfs}
\usepackage{soul}
\usepackage{subfigure}
\usepackage{wrapfig}
\usepackage{xy} 
\usepackage{xcolor}
\usepackage{amsmath,amssymb,amsfonts,amsthm}
\usepackage[ascii]{inputenc}
\usepackage{makecell}
\usepackage[mathscr]{eucal}
\usepackage{cite}
\usepackage{textcomp}
\usepackage{bbold}
\usepackage{ulem}

\allowdisplaybreaks
\renewcommand{\theequation}{\thesection.\arabic{equation}}

\definecolor{dgreen}{rgb}{0,0.7,0}

\graphicspath{{images/}{../Figures}}

\begin{document}

\title{Lax random matrices from Calogero systems}

\author{Jitendra Kethepalli}
\address{International Centre for Theoretical Sciences, Tata Institute of Fundamental Research, Bengaluru -- 560089, India}

\author{Manas Kulkarni}
\address{International Centre for Theoretical Sciences, Tata Institute of Fundamental Research, Bengaluru -- 560089, India}

\author{Anupam Kundu}
\address{International Centre for Theoretical Sciences, Tata Institute of Fundamental Research, Bengaluru -- 560089, India}

\author{Herbert Spohn}
\address{Mathematik Department and Physik Department, Technische Universit{\"a}t\\ M{\"u}nchen, Garching 85748, Germany}

\begin{abstract}
We study a class of random matrices arising from the Lax matrix structure of classical integrable systems, particularly the Calogero family of models. Our focus is the density of eigenvalues for these random matrices. The problem can be mapped to analyzing the density of eigenvalues for generalized versions of conventional random matrix ensembles, including a modified form of the log-gas. The mapping comes from the underlying integrable structure of these models. Such deep connection is confirmed by extensive Monte-Carlo simulations. Thereby we move forward not only in terms of understanding such class of random matrices arising from integrable many-body systems, but also by providing a building block for the generalized hydrodynamic description of integrable systems.
\end{abstract}

\date{\today}
\maketitle

\section{Introduction}
\label{sec1}
\setcounter{equation}{0}
Integrable many-body systems are characterized by an extensive number of conserved charges and have long been a source of fascination for mathematicians and physicists. Their inherent mathematical structure provides a unique avenue for deeper analytical understanding. Simple examples of integrable systems are non-interacting particles and harmonic chains. In the former, the energy of each particle is conserved while in the latter the energy associated with normal modes is conserved, thereby resulting in an extensive number of conserved quantities. When interactions or non-linearities are introduced, mixing among these non-interacting components causes the loss of essentially all local conservation laws leading to non-integrable chaotic systems~\cite{arnold, kam}. However, upon careful fine-tuning, the system may maintain the large number of conserved charges \cite{olshanetsky1981classical,babelon2003introduction,cao1990classical,doyon2017dynamics}. Notable examples include hard rod systems \cite{tonks1936complete,percus1969exact,boldrighini1983one}, the Toda chain \cite{toda1967vibration,  ford1973integrability, flaschka1975, toda1989theory, toda1989studies, spohn2020generalized, spohn2021hydrodynamic} and Calogero models \cite{polychronakos1992, polychronakos2006physics, kulkarni2017emergence, gon2019duality, calogero1969solution, calogero1969ground, calogero1971solution, sutherland1971exact, sutherland1972exact, sutherland1975exact, abanov2011soliton}. 
Somewhat apart is the family of integrable classical spin models, as the Ishimori chain \cite{ishimori, haldane} and the Abolwitz Ladik chain \cite{spohn2022, ALbook, brollo2024}.

For classical systems, establishing the existence of an extensive number of conservation laws is mostly based on the availability of a Lax pair, $L$ and $M$ \cite{olshanetsky1981classical}. These are $N \times N$ matrices whose entries depend on positions $\{q_i\}$ and momenta $\{p_i\}$ of $N$ particles.  For such integrable systems Newton's equations of motion can be recast as 
\begin{align}
 \dot{L} = [L, M],\label{Ldot}
\end{align}
where $[A, B] = AB-BA$ denotes the commutator. From Eq.~(\ref{Ldot}), one deduces that the eigenvalues $\{\lambda_i\}$ of $L$ are time-independent~\cite{olshanetsky1981classical}. However eigenvalues as a function of $\{q_i\}$, $\{p_i\}$ are non-local. For the purpose of a hydrodynamic description 
charges must have a local density. This is achieved by defining the charges $Q_k = \mathrm{tr}[L^k]$, $k=1,...,N$, where $\rm{tr}[A]$ denotes the trace of the matrix $A$. These charges are conserved and have a quasi-local density. The Lax formalism is a crucial tool not only for establishing an extensive number of locally conserved charges but also to provide an algebraic method to handle them.

Recently, there has been a growing interest in the hydrodynamics of integrable systems both theoretically \cite{doyon2020lecture, essler2022short, castro2016emergent, bertini2016transport, bouchoule2022generalized, doyon2023generalized, doyon2018geometric} and experimentally \cite{malvania2021generalized, schemmer2019generalized}. For classical integrable systems, a central tool is the empirical density of eigenvalues (density of states, DOS) of the Lax matrix. The DOS is defined as 
\begin{align}\label{LaxDOS0}
\varrho_N(\lambda) = \frac{1}{N}\sum_{i=1}^N \delta(\lambda - \lambda_i),
\end{align}
where the $\{\lambda_i\}$ are the eigenvalues of the Lax matrix. If positions and momenta are randomly distributed, then $L$ becomes a random matrix. Hence its eigenvalues, $\{\lambda_i\}$, are random and $\varrho_N(\lambda)$ turns into a random 
probability measure. A specific example is the case where the positions and the momentum are distributed according to the thermal Gibbs-Boltzmann distribution 
\begin{align}
P(\{q_i, p_i\}_{i=1}^N)= \frac{1}{Z} \exp\left(-\beta H(\{q_i, p_i\})\right),
\end{align}
at inverse temperature  $\beta$, where $Z$ is the normalizing partition function. When these positions and momentum are entries of the Lax matrix then we refer to such density of eigenvalues as the thermal Lax DOS. In the context of a hydrodynamic description of integrable systems, the Generalized Gibbs ensemble (GGE) is of central importance. In this case, one expects that  
\begin{align}\label{LaxDOS}
\lim_{N \to \infty}\varrho_N(\lambda) = \varrho(\lambda)
\end{align}
with probability one in the thermodynamic limit. The limit density, $\varrho(\lambda)$, is deterministic. 

To compute a limiting  DOS is an extensively studied problem in the context of Random Matrix Theory, starting with Wigner ensembles~\cite{wigner1951}. In the same spirit, it is of interest to study the DOS of eigenvalues associated with the Lax matrix $L$ \cite{bogomolny2009random} from integrable classical Hamiltonians. For short, we call them \textit{random Lax matrices}. They are part of classical integrable many-body theory and constitute an interesting novel class of random matrices. 

In this article, we consider the hyperbolic Calogero model, which consists of $N$ particles interacting via a short-range repulsive pair-wise potential.  The Hamiltonian reads \cite{olshanetsky1981classical}
\begin{align}\label{hcm: PE}
   H_{\mathrm{c}} \big(\{q_i, p_i\}\big) = \sum_{i=1}^{N}\tfrac{1}{2} p_i^2 +  \sum_{1\leq i < j}^N V_{\mathrm{c}}(q_i-q_j), \quad  \quad V_{\mathrm{c}}(r) = \frac{1}{4\sinh^2\left(r/2\right)}.
\end{align}
Here $q_i$ and $p_i$ denote the position and momentum of the $i$-th particle, respectively, with $i=1,2,...,N$.  The Hamiltonian $H_{\mathrm{c}}$ models a 
one-dimensional fluid. Only for the very specific choice of $V_{\mathrm{c}}$ in Eq.~\eqref{hcm: PE} the fluid is integrable. Therefore we refer to $H_{\mathrm{c}}$ as Calogero fluid~\cite{olshanetsky1981classical}. The associated Lax pair, $L_{\mathrm{c}}$ and $M_{\mathrm{c}}$, are $N\times N$ matrices  explicitly given by
\begin{align}
\begin{split}
 [L_{\mathrm{c}}]_{ij} &= \delta_{ij}p_j + \mathrm{i}(1- \delta_{ij})\frac{1}{2\sinh\left(\tfrac{1}{2}\left(q_i- q_j\right)\right)} ,\\
[M_{\mathrm{c}}]_{ij}& =  \mathrm{i}\delta_{ij} \sum_{k = 1,k\neq j}^N \frac{1}{4\sinh^2\left(\tfrac{1}{2}\left(q_j- q_k\right)\right)}  - \mathrm{i}(1- \delta_{ij})\frac{ \cosh(\tfrac{1}{2}(q_i- q_j))}{4\sinh^2\left(\tfrac{1}{2}(q_i- q_j)\right)}.
   \end{split}
   \label{Lax: hcm}
\end{align}
Here the subscript ``c'' in Eq.~\eqref{hcm: PE} and~\eqref{Lax: hcm} stands for the Calogero fluid.

The central agenda of this article is to study the DOS of the Lax matrix $L_{\mathrm{c}}$ in case the system is in Gibbs thermal equilibrium. For this purpose one has to confine a system of $N$ particles to a region approximately of size $\ell$ and then take the limit $N,\ell \to \infty$ at fixed particle density $\bar{\rho}=N/\ell$. The asymptotic DOS is expected to be independent of the precise confining mechanism. As discussed in \cite{spohn2023hydrodynamic}, an analytically very powerful choice is the fine-tuned external potential 
\begin{align}\label{vex}
 U_\mathrm{C}(\{q_i\}) = \sum_{i=1}^N e^{-\ell/2} \mathrm{\cosh}(q_i).
\end{align}
Here the subscript ``C'' stands for the particular cosh form of the confining potential. The parameter $\ell$ is the size of the confinement. In this very specific case, a closed expression for the joint distribution of the $N$ eigenvalues is available~\cite{spohn2023hydrodynamic}. We call such a choice as the cosh-confined  Calogero fluid. Note that throughout the manuscript, we use small alphabets for the type of interaction and capital alphabets to represent the nature of the external trap in the subscripts.

Another simple way to confine particles is to place them inside a box with hard walls at $\pm\ell/2$, which amounts to the external potential 
\begin{align}\label{vbox}
    U_{\mathrm{B}}(x) =\begin{cases}
        0 &~\text{for}~~|x|<\tfrac{1}{2}\ell,\\
        \infty &~\text{for}~~|x|\geq \tfrac{1}{2} \ell.
    \end{cases}
\end{align}
The subscript ``B'' labels the box potential.

An alternative choice to confine particles is to place them on a ring of circumference $\ell$. Since the interaction potential $V_\mathrm{c}$ is of the infinite range one has to include all image positions, which means to periodize the
potential. The periodized interaction potential can still be obtained explicitly by using the identity
\begin{align}\label{sinhtowp}
    \wp\bigg(r \bigg|\frac{\ell}{2}, \mathrm{i} \pi \bigg)-{\wp\bigg(\frac{\ell}{2}\bigg|\frac{\ell}{2}, \mathrm{i} \pi \bigg)} = \sum_{n\in\mathbb{Z}}\left[\frac{1}{4\sinh^{2}\left(\frac{r}{2} + n\ell\right)}-{\frac{1}{4\sinh^{2}\left(\frac{\ell}{4} + n\ell\right)}}\right].
\end{align}
Here $\wp$ denotes the double periodic Weierstrass function. The Hamiltonian for the periodized Calogero fluid is then given by
\begin{align}\label{weier: PE}
    H_{\mathrm{w}}\big(\{q_i, p_i\}\big) = \sum_{i=1}^{N}\tfrac{1}{2}p_i^2  +  \sum_{1 \leq i <j}^N V_{\mathrm{w}}(q_i-q_j)
\end{align}
with interaction potential
\begin{align}\label{weier: interaction}
    V_{\mathrm{w}}(r) = \wp\bigg(r \bigg|\frac{\ell}{2}, \mathrm{i} \pi \bigg)-{\wp\bigg(\frac{\ell}{2}\bigg|\frac{\ell}{2}, \mathrm{i} \pi \bigg)}.
\end{align}
The subscript ``w'' is a label for the Weierstrass function appearing in the formula. Note that any additional constant to the Hamiltonian in Eq.~\eqref{weier: PE} $H_{\mathrm{w}} \to H_{\mathrm{w}} + \mathrm{Constant}$ does not modify the Newtons equation of motion. In the literature, the model is known as the elliptic Calogero model  \cite{calogero2014}.  This model still possesses the Lax matrix structure with the following Lax pair 
\begin{align}
\begin{split}
    [L_{\mathrm{w}}]_{ij} &= \delta_{ij}p_j + (1- \delta_{ij}){\alpha_{\mathrm{w}}(q_i-q_j)}, \\
[M_{\mathrm{w}}]_{ij}& =  \mathrm{i}\delta_{ij}\sum_{\substack{k=1 \\ k\neq i}}^N {V_{\mathrm{w}}(q_k-q_j) +(1- \delta_{ij}) \alpha_{\mathrm{w}}'(q_i-q_j)}
\end{split}
\label{Lax: weier}
\end{align}
where $\alpha_{\mathrm{w}}'(r) = \tfrac{d}{dr}\alpha_{\mathrm{w}}(r)$ and
\begin{align}
    \alpha_{\mathrm{w}}(r) &= \mathrm{i} \frac{\sqrt{e_1-e_3}~\mathrm{Cn}(r\sqrt{e_1-e_3}|m)}{\mathrm{Sn}(r\sqrt{e_1-e_3}|m)},~~\text{with}~~m^2 = \frac{e_2-e_3}{e_1-e_3}~~\text{where}\\
    e_1 &= \wp\bigg(\frac{\ell}{2} \bigg|\frac{\ell}{2}, \mathrm{i} \pi \bigg), ~ e_2 = \wp\bigg(-\frac{\ell}{2}-\mathrm{i} \pi\bigg|\frac{\ell}{2}, \mathrm{i} \pi \bigg), ~\text{and}~    e_3 = \wp\bigg(\mathrm{i} \pi \bigg|\frac{\ell}{2}, \mathrm{i} \pi \bigg).
\end{align}
Here the Jacobi Sin and Jacobi Cos functions, $\mathrm{Sn}(r)$ and $\mathrm{Cn}(r)$, are given in Eq.~\eqref{app:jacobi}. Note that $\alpha_{\mathrm{w}}(r)$ is a odd function and it satisfies the relation $V_{\mathrm{w}}(r) = -[\alpha_{\mathrm{w}}(r)]^2$. In the \ref{applaxwiers}, we describe further details and general properties of the Lax matrix $L$ and $M$ and the relation between the Weierstrass function and the Jacobi Sin function. 

In our article, we numerically investigate the Lax DOS in Gibbs thermal equilibrium (thermal Lax DOS) of the Calogero fluid 
for a wide range of densities and temperatures. We also elucidate the dependence on 
boundary conditions.
Our main contributions are as follows:
\begin{enumerate}
    \item[(i)] Using the Monte Carlo simulations we compute the thermal Lax DOS of the Calogero fluid for three boundary conditions (a) the cosh confining trap, (b) the confining box and (c) on a ring. For the case of cosh confinement we compute the Lax DOS using two methods: by direct diagonalization and based on the exact joint distribution of eigenvalues, see~\cite{spohn2023hydrodynamic}. For the other two cases (b) and (c) we use the method based on direct diagonalization of the Lax Matrix.
     \item[(ii)] To elucidate the role of the boundary conditions, we compare the Lax DOS of the Calogero fluid for different boundary conditions.
    \item[(iii)] We demonstrate that, at low density the Calogero fluid is well approximated by the Toda chain and at high density by the rational Calogero model. For these parameters we compare the DOS of the Calogero fluid with a direct numerical simulation of the approximating models in different boundary conditions.
\end{enumerate}
 
The article is organized as follows. In Section~\ref{sec:blg} we will present some details of the Calogero fluid and review the derivation of the joint distribution of the Lax matrix eigenvalues. In Section~\ref{sec: num}, we present our numerical findings on the Calogero fluid. Section~\ref{sec:spl} is devoted to low and high density limits of the Calogero fluid. We summarize our results along with an outlook in Section~\ref{sec:conc}. Some details are delegated to the Appendix.

\section{Eigenvalues of the random Lax matrix}
\label{sec:blg}
\setcounter{equation}{0}
We briefly recap the derivation of the joint distribution of eigenvalues of the Lax matrix for the Calogero fluid \cite{spohn2023hydrodynamic}. A key step is to canonically transform from positions on the real line and their momenta to scattering coordinates, i.e. to the asymptotic momenta $\{\lambda_i\}$ and their associated scattering shifts $\{\phi_i\}$. Then the dependence on the eigenvalues trivializes. Miraculously, for the choice  Eq.~\eqref{vex} of the confining potential, the integration over the scattering shifts can still be performed. 

In more detail, the configuration of positions and momenta are chosen from the distribution, see Ref.~\cite{spohn2021hydrodynamic}, 
\begin{align}\label{P(q_i,p_i)}
P(\{q_i, p_i\};\ell,\beta)~d^Npd^Nq = \frac{1}{Z(N,\ell,\beta)}\exp\left(-\tfrac{1}{2}\beta\mathrm{tr}[L^2]-U_\mathrm{C}(\{q_j\}) \right)~d^Npd^Nq,
\end{align}
where $U_{\mathrm{C}}\left(\{q_i\}\right)$ is given in Eq.~\eqref{vex}, the momenta $p_i \in \mathbb{R}$, the positions are ordered  as $-\infty < q_1 <\ldots < q_N < \infty$, $d^Npd^Nq=\prod_{i=1}^Ndq_idp_i$, and $Z(N,\ell,\beta)$ is the normalizing partition function. 

In terms of the scattering coordinates, $\{\lambda_i, \phi_i\}$, the external potential has the simple form
\begin{align}
U_{\mathrm{C}}(\{q_i\}) \mapsto  \sum_{i=1}^Ne^{-\frac{\ell}{2}}Y_i~\text{cosh}(\phi_i), \label{pot-in-sc-co}
\end{align}
where
\begin{align}
Y_i&=\prod_{m=1, m\neq i}^N\Big(1 +\frac{1}{(\lambda_m -\lambda_i)^2}\Big)^{1/2}. \label{Y_i}
\end{align}
Since $\{q_i,p_i\} \mapsto \{\lambda_i,\phi_i\}$ is a canonical transformation, {\it i.e.} it satisfies 
$d^Np\,d^Nq = d^N\lambda\, d^N\phi=\prod_{i=1}^N d\lambda_i\,d\phi_i$. Consequently the joint distribution Eq.~\eqref{P(q_i,p_i)} is transformed to 
\begin{align}\label{poflambdaphi}
P(\{\lambda_i, \phi_i\};\ell,\beta)d^N\lambda\, d^N\phi = 
\frac{1}{N!Z(N,\ell,\beta)}\exp\left(\!\!-\tfrac{\beta}{2}\sum_{i=1}^N \!\! \lambda_i^2 -\sum_{i=1}^N \!\! e^{-\frac{\ell}{2}}Y_i \text{cosh}(\phi_i) \right)\!\!d^N\lambda\, d^N\phi.
\end{align}
Integrating over $\{\phi_i\}$ results in the following joint distribution for the eigenvalues
\begin{align}\label{poflambda0}
\mathcal{P}(\{\lambda_i\};\ell,\beta) & =\frac{1}{N!~Z(N,\ell,\beta)}  
\exp\left(-H_\mathrm{b}\left(\{\lambda_i\};  \ell,\beta\right)\right),
\end{align}
where
\begin{align}\label{poflambda1}
H_\mathrm{b}\left(\{\lambda_i\};  \ell,\beta\right)= \tfrac{\beta}{2}\sum_{i=1}^N\lambda_i^2  - \sum_{i=1}^N \log\Bigg(2K_0\bigg( 2 \exp\Big[ - \tfrac{\ell}{2} + \log Y_i\Big]\bigg)\Bigg).
\end{align}
Here $K_0(x)$ is the zeroth order modified Bessel function of the second kind defined as
\begin{align}\label{besselk}
K_0(x) = \int_0^\infty \mathrm{d}t\exp(- x \cosh{t}). 
\end{align}
We use the subscript ``b'' in Eq.~\eqref{poflambda1} to remind on the presence of this Bessel function.
The asymptotics of the Bessel  function are $K_0(x) = - \log x$ for $x \to 0$ and $ K_0(x) = (\pi/2x)^{1/2}\mathrm{e}^{-x}$ for $x \to \infty$. 

For large $N$ the lower tail dominates and, in this approximation,  $H_\mathrm{b}$ can be written as a functional of the asymptotic Lax DOS which is given by 
\begin{align}\label{frho}
\mathcal{F}_{\mathrm{c}}[\varrho] =  
 N\int_\mathbb{R}\mathrm{d}w\varrho(w)\Big(\tfrac{\beta}{2}w^2 -1  + \log \varrho(w) - \log\big(\nu +  \int_\mathbb{R}\mathrm{d}w'\varrho(w')\phi(w - w')\big) \Big). 
\end{align}
This formula holds in case $\nu + \int_\mathbb{R}\mathrm{d}w'\varrho(w')\phi(w - w') >0$. Otherwise $\mathcal{F}_{\mathrm{c}}[\varrho] = \infty$.
Here   $\nu = \tfrac{1}{\bar{\rho}} = \ell/N$ is the inverse density and $\phi(w - w')$ is the two-particle  Calogero scattering shift, 
\begin{align}\label{eq:phi}
\phi(w) = -\ln\left(1+\frac{1}{w^2}\right).
\end{align}
In Eq.~\eqref{frho} the first and last term are effective energies. The entropy, $-\int_\mathbb{R}dw~\varrho(w) \log \varrho(w)$, results from the a priori measure. $\mathcal{F}_{\mathrm{c}}[\varrho]$ has to be minimized over all $\varrho(w) >0$ normalized as 
$\int_\mathbb{R} \mathrm{d}w \varrho(w) = 1$.  The unique minimizer of the free energy functional $\mathcal{F}_{\mathrm{c}}[\varrho]$ is the Lax DOS of the Calogero fluid in the thermodynamic limit.\\\\
\textit{Remark}. Clearly, $H_\mathrm{c} = \tfrac{1}{2}\mathrm{tr}[L_\mathrm{c}^2]$,
which yields the sum over $\lambda_i^2$
in the spectral representation for the thermal state. For a general GGE one would have to replace the quadratic function 
by some arbitrary $h(\lambda_i)$. To have a well-defined
partition function $h$ has to increase to infinity on both sides. With this modification, Eq.~\eqref{poflambda0} 
is the joint distribution of eigenvalues for a general GGE.

\section{Numerical study of the Lax DOS}
\label{sec: num}
\setcounter{equation}{0}
We present the numerical results for the thermal Lax density of states (DOS) of the Calogero fluid. Since the interaction potential is short-ranged, the equilibrium state has a finite correlation length and one expects the DOS to be independent of boundary conditions. To obtain a global overview of the DOS as a function of the particle density $\bar \rho$ and the temperature $T =\beta^{-1}$, we first use the box potential of length $\ell$ given in Eq.~\eqref{vbox}. Then the Boltzmann weight is given by
\begin{equation}\label{B}
\exp\big(-\beta H_{\mathrm{c}}\left(\{q_i, p_i\}\right) -  U_\mathrm{B}\left(\{q_i\}\right)\big).
\end{equation}
The configurations $\{q_i,p_i\}$ are sampled by a Monte Carlo (MC) algorithm. For each configuration the corresponding Lax matrix $L_\mathrm{c}$ is diagonalized, thereby obtaining the DOS. Interestingly, specific limits of the Calogero fluid sill define integrable many-particle models, on which we elaborate below.

When the particle density is low, $\bar \rho \ll 1$, then the typical gap $\nu = 1/\bar \rho$ between particles is large, $\nu \gg 1$. At this scale, the interaction potential of the Calogero fluid approximately behaves as an exponentially decaying function of the interparticle distance $r$. Furthermore, the nearest-neighbor interaction exponentially dominates the next nearest-neighbor one. Thereby one arrives at the Toda chain for which particles with neighbouring index interact through a potential of the form
\begin{align}\label{Vtoda}
    V_{\mathrm{to}}(r) = \exp(-r).
\end{align}
Here the subscript ``to'' stands for Toda chain. The Toda chain is an integrable system \cite{toda1967vibration, olshanetsky1981classical} and its Lax DOS has been studied when the system is confined to an external trap ~\cite{spohn2023hydrodynamic}. Therefore, at a given low particle density the Toda chain and the Calogero fluid are expected to display similar equilibrium properties like Lax DOS. We provide a brief discussion of the Lax DOS for the Toda chain  in \ref{ldosptoda}.

On the other hand, when the density of particles is high, $\bar \rho \gg 1$, particles are very close. At this scale, the interaction potential of the Calogero fluid behaves as an inverse quadratic function of interparticle distance. This is the interaction potential of the rational Calogero model where the particles repel each other with a power-law potential of the form
\begin{align}\label{Vrational}
    V_{\mathrm{r}}(r) = \frac{1}{r^2}.
\end{align}
Here the subscript ``r'' stands for rational Calogero model. The rational Calogero model is also an integrable system  \cite{calogero1969solution, polychronakos2006physics, olshanetsky1981classical}. When these particles are confined to a ring of length $\ell$, the periodized interaction Eq.~\eqref{Vrational} results in the following pair potential
\begin{align}
    V_{\mathrm{t}}(r) = \frac{1}{\sin^2\left(\frac{\pi}{\ell}r\right)},
\end{align}
which is known  trigonometric Calogero model \cite{choquardclassical, olshanetsky1981classical}.  Here the subscript ``t'' refers to the trigonometric form of the interaction. This model is also integrable. The Lax DOS of the trigonometric Calogero model is well understood analytically using the Thermodynamic Bethe ansatz (TBA)~\cite{spohn2023hydrodynamic}. We briefly discuss the Lax DOS for this model in \ref{trig}. The Lax DOS of the Toda chain and the trigonometric Calogero model will be used to benchmark our findings for the Lax DOS of the Calogero fluid. 

\begin{figure}
\hspace{-1cm}\includegraphics[scale=0.5]{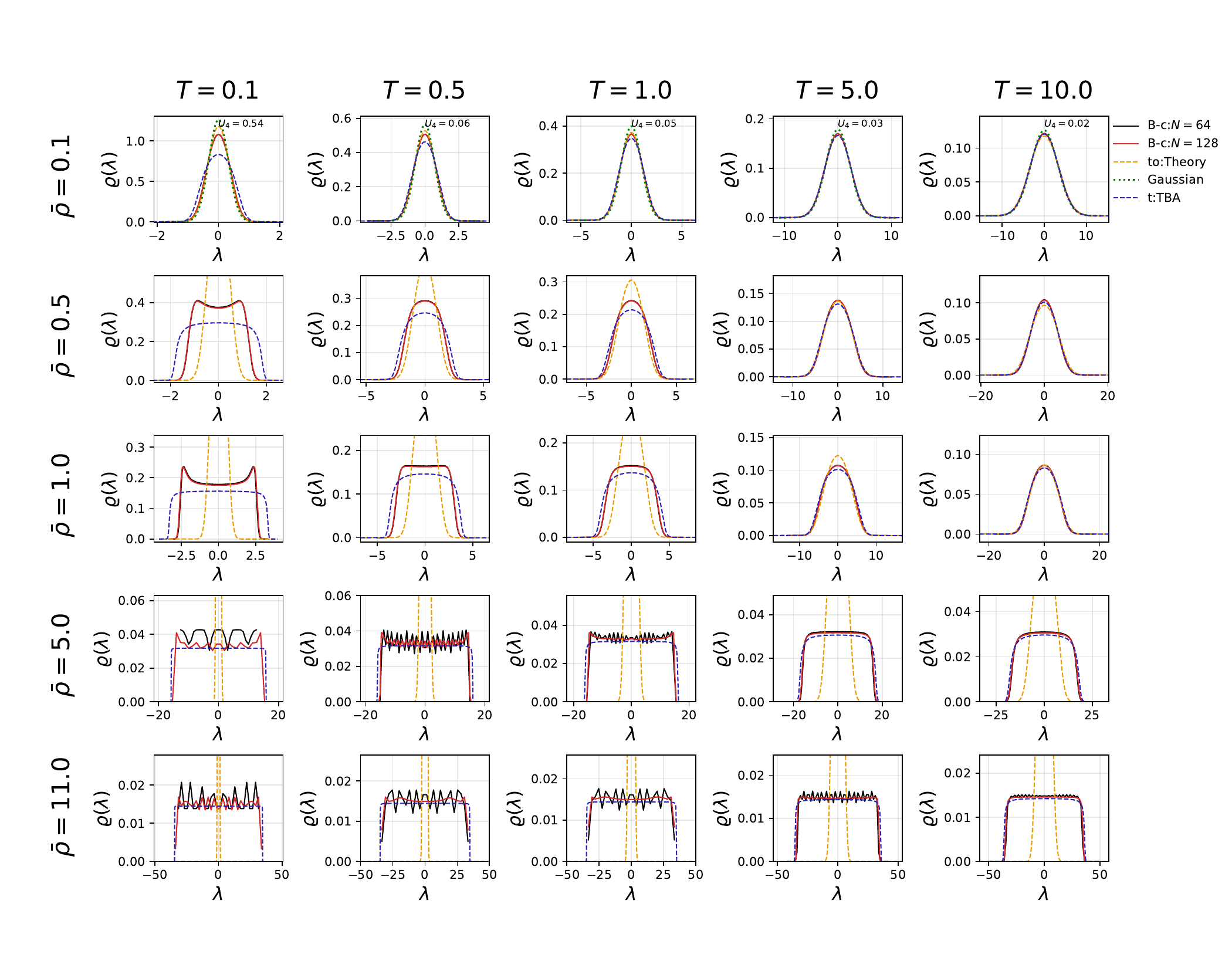}
    \caption{The plot shows the thermal Lax DOS of the Calogero fluid in a box for $N = 64$ and $ 128$ (black and red solid lines) at temperatures $T = 0.1, 0.5, 1.0, 5.0$ and $10.0$ and particle densities $\bar \rho = 0.1, 0.5, 1.0, 5.0$ and $11.0$. The average is over $10^5$ samples in every subplot. We compare the Lax DOS of this model with the analytical expressions of Lax DOS of the Toda chain (dashed yellow line)  and the tignometric Calogeoro model  (dashed blue line) \cite{spohn2023hydrodynamic} provided in ~\ref{ldosptoda} and \ref{trig}  respectively. For some parameter regimes (relatively low density and high temperatures) the profiles closely resemble a Gaussian form. To quantify this proximity, we give the value of the Binder cumulant, see Eq.~\eqref{binder-cumulant}, of some profiles.
    Furthermore, for a visual comparison, we also plot a Gaussian distribution (dotted green line) with a variance equal to the temperature $T$ in the first row.}
    \label{fig:dos_vs_T_rbar}
\end{figure}

The MC results of Lax DOS of Calogero fluid in box confinement are displayed in Fig. \ref{fig:dos_vs_T_rbar} where we present the DOS for different values of particle density $\bar{\rho}$ and temperature $T$. We observe that for relatively low densities of particles the Lax DOS of the Calogero fluid closely resembles that of the Toda chain whilst at relativey high densities, the resemblance is closer to that of Trignometric Calogero model. At low density and high temperature $T$, we find that the DOS is well approximated by a Gaussian with variance  equal to $T$. To check the Gaussianity of the DOS at low density and high temperature, we compute the Binder cumulant ($U_4$), defined as
\begin{align}\label{binder-cumulant}
    U_4 = 1 -\frac{\langle \lambda^4 \rangle}{3\langle \lambda^2 \rangle^2},
\end{align}
 where $\langle \lambda^n \rangle = \int_{-\infty}^\infty d \lambda ~\lambda^n \varrho(\lambda)$. For a Gaussian distribution $U_4 = 0$, whereas $U_4 \neq 0$ indicates a deviation from it. 
 
As the density of particles is increased, the DOS starts deviating from Gaussian even at higher temperatures. While at low temperatures, the repulsive interaction first flattens the top and even produces a dip at the symmetry axis. At the highest densities, at any temperature, the DOS tries to be flat except very close to the edges of the profile.  

\begin{figure}[tbh]
\centering
\includegraphics[scale=0.6]{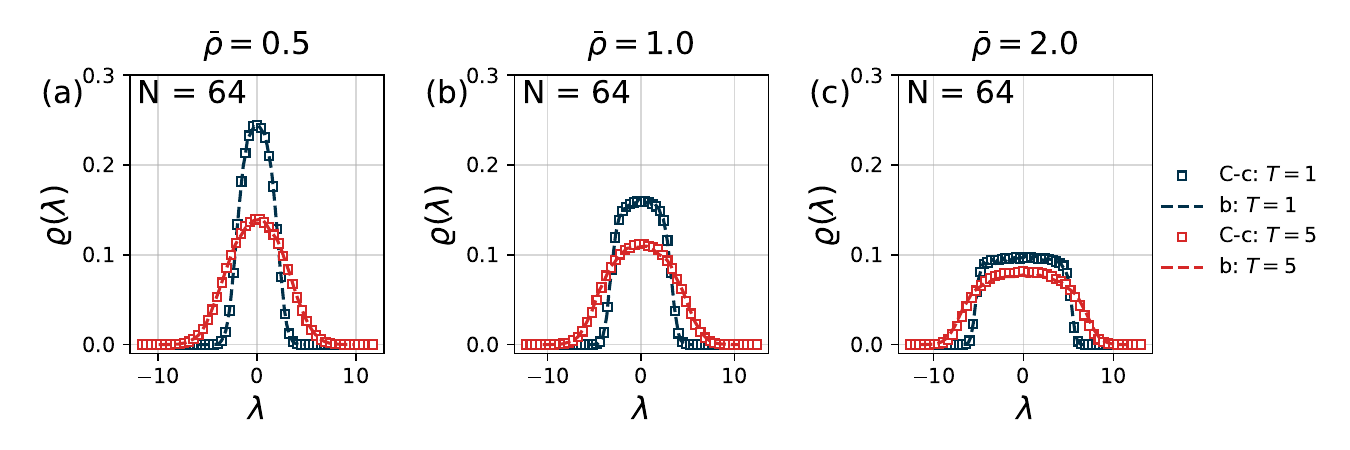}
\caption{The plot shows the thermal Lax DOS for $N=64$ at temperatures $T=1,5$ and particle densities $\bar{\rho} = 0.5,1,2$. The average is over $10^5$ samples for every plot. We compare the Calogero fluid with external cosh potential (symbol C-c) with MC of the exact probability density of eigenvalues (symbol b) according to Eq.~\eqref{poflambda0}.}
\label{fig:comparbessinh}
\end{figure}

Our next goal is to investigate the dependence of the Lax DOS on the choice of the boundary conditions in the above discussed models. We start with the Calogero fluid in cosh confinement. For this external potential, there is an explicit formula for the joint distribution of eigenvalues in Eq.~\eqref{poflambda0}.  
We first numerically demonstrate agreement between DOS obtained using MC for the Calogero energy function in cosh confinement and the direct MC simulation of the modified log-gas with energy function given in Eq.~\eqref{poflambda1}. In Fig.~\ref{fig:comparbessinh}, we plot the corresponding DOS for $\bar{\rho} = 0.5,1,2$ and temperatures $T=1,5$ at system size of $N=64$. We find excellent agreement between the two methods in all cases. Thus an alternative approach to get Lax DOS is to directly MC sample the distribution in Eq.~\eqref{poflambda0}. Somewhat unexpectedly, this route turns out to be considerably slower numerically as the computation of the Bessel $K$ function is numerically expensive.   Hence, for further investigation on this model we stick to the MC simulation with the Boltzmann weight $\exp(-\beta H_{\rm c}(\{q_i,p_i\})-V_{\mathrm{C}}(\{q_i\})$ where $ H_{\rm c}(\{q_i,p_o\})$ and $V_{\mathrm{C}}(\{q_i\})$ are given in Eq.~\eqref{hcm: PE} and~\eqref{vex}.  

\begin{figure}[htb]
\centering
\includegraphics[scale=0.62]{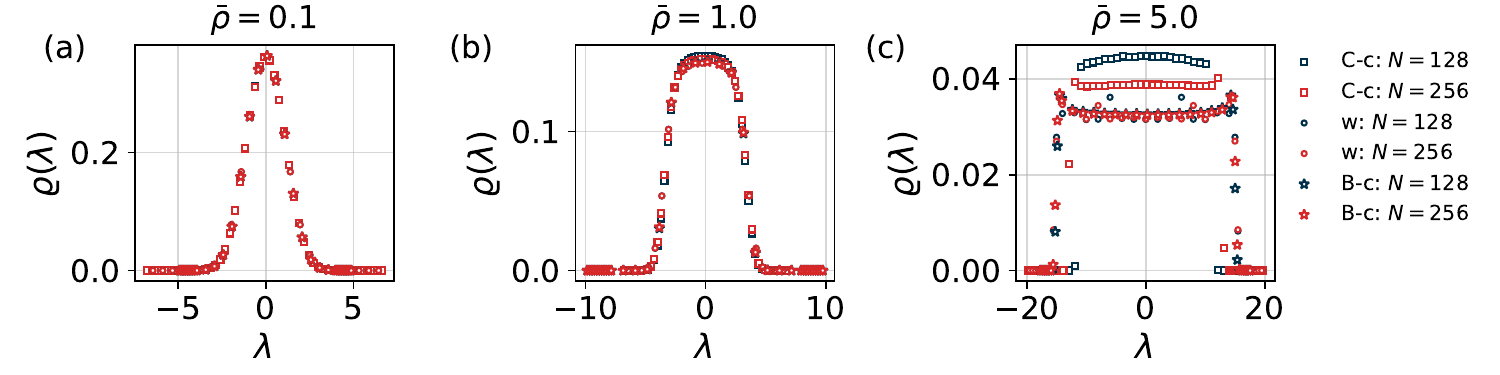}
\caption{The plot shows the thermal Lax DOS for $N=128, 256$ at temperature $T=1$  and particle densities $\bar{\rho} = 0.1, 1.0, 5.0$. The average is taken over $10^5$ samples for each plot. We compare the Calogero fluid with external cosh potential ( square symbol C-c), hard box potential (star symbol B-c), and periodic boundary conditions (circle symbol w).}
\label{fig:comparsinhweier}
\end{figure}

To compute the Lax DOS of the Calogero fluid for the periodic boundary conditions, we perform MC simulations with the Boltzmann weight 
\begin{equation}
\exp\big(-\beta H_{\mathrm{w}}\left(\{q_i, p_i\}\right)\big),
\end{equation}
where $H_{\rm w}(\{q_i, p_i\})$ given in Eq.~\eqref{weier: PE} corresponds to the energy function of the periodized Calogero fluid {\it i.e.} the elliptic Calogero model.
For our choice of parameters, the double periodic Weierstrass function appearing in Eq.~\eqref{weier: PE}, is approximated by truncating the series in Eq.~\eqref{sinhtowp} at $n=\pm2$. This approximation works well when $N$ and $\ell$ are large.

In Fig.~\ref{fig:comparsinhweier} we compare the Lax DOS of Calogero fluid corresponding to fixed values of the parameters ($\bar{\rho}$ and $T$) for different boundary conditions --- cosh potential, hard box potential, and on a ring. In particular, for each case, we plot DOS profiles for three values of $\bar{\rho} = 0.1, 1.0, 5.0$ at temperature $T=1$ for system sizes $N=128, 256, 512$. We observe that with increasing $N$ the Lax DOS for different boundary conditions agree, which confirms the common expectation that the details of the boundary conditions do not matter for thermodynamic quantities. However, while there is a good agreement for low densities, for higher densities there are visible differences. This will be discussed in the next section, where we also discuss the slow convergence of the bulk Lax DOS when increasing the density of particles, $\bar{\rho}$.

\section{Low- and high-density limit}
\label{sec:spl}
\setcounter{equation}{0}
We discuss the Lax DOS for the Calogero fluid in the limiting cases when the particle density $\bar{\rho}$ is either low or high. As can be seen from Fig. \ref{fig:dos_vs_T_rbar}, the behaviour is very different. For low density, particles are far apart and the interaction potential can be approximated by $ V_{\mathrm{c}}(r) \simeq V_{\rm to}(r)= \exp(-r)$, $r\gg1 $, see Eq.~\eqref{Vtoda}. In addition, the next nearest neighbour interaction is exponentially smaller compared to the nearest neighbour one, which yields the nearest neighbour chain with the exponential interaction Eq.~\eqref{Vtoda}. In the opposite limit, one arrives at $V_{\mathrm{c}}(r) \simeq V_{\rm r}(r)= r^{-2}$, $r\ll 1$, see Eq.~\eqref{Vrational}. The interaction is still pair-wise, thereby yielding the rational Calogero model with the interaction potential given by Eq.~\eqref{Vrational}. Both the Toda chain~\cite{toda1967vibration, toda1989theory, toda1989studies, ford1973integrability, flaschka1975} and the rational Calogero model~\cite{calogero1969solution, calogero1969ground, calogero1971solution, sutherland1971exact, sutherland1972exact, sutherland1975exact, abanov2011soliton} are integrable systems and have Lax pairs. Thus, beyond the mere asymptotics of the interaction potential $V_{\mathrm{c}}$, one expects that even the DOS of the Calogero fluid is well approximated by the DOS of the limiting cases. The range over which such approximations are valid can be found out only numerically, as for example in  Fig.~\ref{fig:dos_vs_T_rbar}.

\noindent
\textit{Low density:} In this limit the Calogero Hamiltonian is approximated by the open Toda chain 
\begin{align}\label{toda: PE}
        H_{\mathrm{to}}\left(\{q_i, p_i\}\right) = \sum_{i=1}^{N}\tfrac{1}{2}p_i^2  + \sum_{i=1}^{N-1} \exp\left(-(q_{i+1}-q_i)\right).
\end{align}
This Hamiltonian has the Lax matrix pair  
\begin{align}\label{lax: toda} 
\begin{split}
[L_{\mathrm{to}}]_{jj} &= p_j,\quad j = 1,...,N,  \\ 
[L_{\mathrm{to}}]_{jj+1} &= \exp\left(-\tfrac{1}{2}(q_{j+1}-q_j)\right) = [L_{\mathrm{to}}]_{j+1j}, \quad j=  1,...,N-1, \\ 
[L_{\mathrm{to}}]_{ij} &= 0, \quad \text{otherwise},
\end{split}
\end{align}
and 
\begin{align}\label{M: toda}
\begin{split}
    [M_{\mathrm{to}}]_{jj+1} &= -[M_{\mathrm{to}}]_{j+1j} = \exp\left(-\tfrac{1}{2}\left(q_{j+1}-q_j\right)\right),  \quad j=  1,...,N-1,\\
    [M_{\mathrm{to}}]_{ij} &= 0 , \quad \text{otherwise.}
    \end{split}
\end{align}
In analogy to the Calogero fluid, the Toda chain is confined by an external potential~\cite{spohn2023hydrodynamic}, which is exponential and acts only on the first and $N$-th particle as
\begin{align}\label{vextoda}
    U_\mathrm{E}(q_1, q_N,\ell) = \exp\left(-\left(q_1 + \tfrac{1}{2}\ell\right)\right) + \exp\left(-\left( \tfrac{1}{2}\ell-q_N\right)\right).  
\end{align}
The subscript ``E'' stands for the exponential potential. This special choice of the confinement is well studied in Ref.~\cite{spohn2023hydrodynamic}. At inverse temperature $\beta$, the equilibrium Boltzmann weight for the Toda chain   is given by
\begin{align}
\exp\big(-\beta H_{\mathrm{to}}(\{q_i,p_i\}) - U_\mathrm{E}(q_1, q_N,\ell)\big).
\label{toda: pofqp}
\end{align}
On the other hand, when the Toda particles are confined a box potential, the Boltzmann weight is given by
\begin{align}
\exp\big(-\beta H_{\mathrm{to}}(\{q_i,p_i\}) - \sum_{i=1}^N U_\mathrm{B}(q_i)\big).
\label{toda: pofqpvbox}
\end{align}
In case of periodic boundary conditions, one starts from the Hamiltonian of the closed Toda chain given by 
\begin{align}\label{ptoda: PE}
        H_\mathrm{pt}\left(\{q_i, p_i\}\right) = \sum_{i=1}^{N}\tfrac{1}{2}p_i^2  + \sum_{i=1}^{N} \exp\left(-(q_{i+1}-q_i)\right),
\end{align}
where ``pt'' stands for `periodic Toda' in the sense that the positions of the particles are $q_i~\mathrm{mod}(\ell)$ with $\mathrm{mod}$ being the modulo operation. This system is also integrable \cite{flaschka1975} and its Lax pair, $L_\mathrm{pt}$ and  $M_\mathrm{pt}$,  differs from the open chain
only by the additional matrix elements $[L_\mathrm{pt}]_{1,N}=[L_\mathrm{pt}]_{N,1}= \exp\left(-\tfrac{1}{2}(\ell+q_{1}-q_N)\right)$ and 
$[M_\mathrm{pt}]_{1,N}=-[M_\mathrm{pt}]_{N,1}= -\exp\left(-\tfrac{1}{2}(\ell+q_{1}-q_N)\right)$. However, the Hamiltonian Eq.~\eqref{ptoda: PE} has no confining mechanism, which now is ensured by the boundary condition
\begin{equation}
q_{N+1} = q_1 + \ell.   
\end{equation}
To construct the $L_\mathrm{pt}$ we use the thermal state obtained from MC simulation for $H_\mathrm{pt}$ imposing the above constraint. 

The DOS of the tridiagonal Lax matrix Eq.~\eqref{lax: toda} has been well studied for arbitrary particle densities in the presence of external pressure in Ref.~\cite{spohn2020generalized}. Since the positions
of the Toda particles are not ordered, the DOS at high density has a shape very different from the Calogero fluid. Increasing the pressure, one starts from the single Gaussian peak which then changes to a double hump and,
in the limit of high pressure, takes the form $\rho(w) = \tfrac{1}{\pi\sqrt{1 - w^2}}$ for
$|w| < 1$, $\rho(w) = 0$ for $|w| > 1$, in appropriate units \cite{cao2019}.

\begin{figure}[t]
\centering
\includegraphics[scale=0.64]{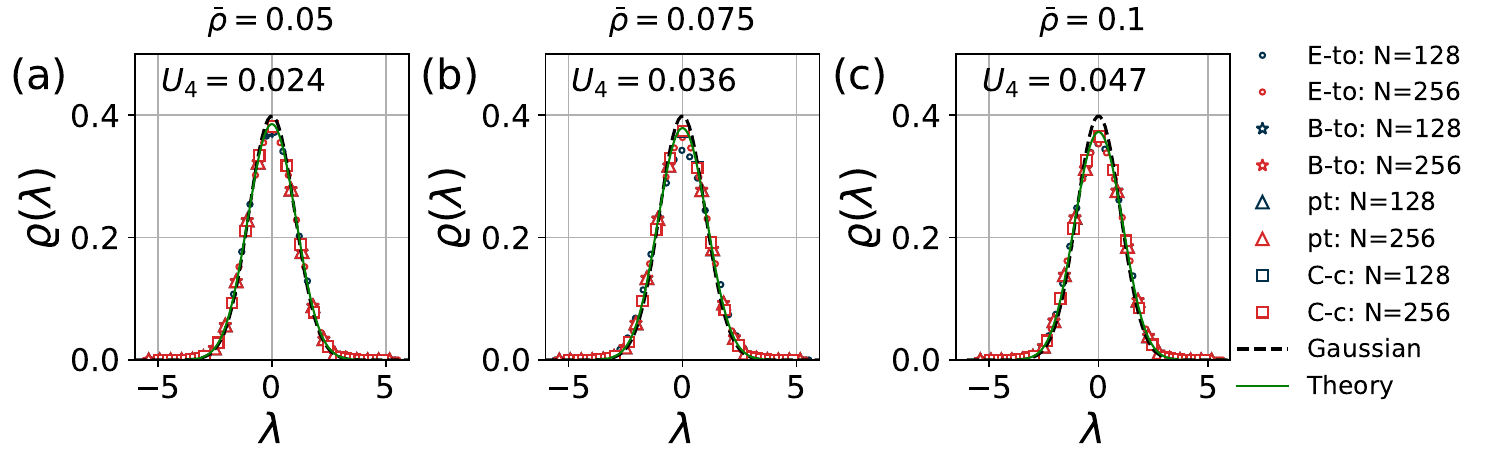}
\caption{The plot shows the thermal Lax DOS for the Toda chain confined to (i) an external trap given by Eq.~\eqref{vextoda} (circle symbol ``E-to''), (ii) a ring {\it i.e.} with periodic boundary conditions (triangle symbol ``pt'') and (iii) a box potential (star symbol ``B-to'') in comparison with the cosh-confined Calogero fluid (square symbol ``C-c''). The number of particles is $N = 128,256$, the temperature $T=1$, and the particle densities in (a), (b), (c) respectively are $\bar{\rho} = 0.05, 0.075, 0.1$  which are considered to be relatively small. The green solid line is the infinite volume solution provided in Eq.~\eqref{dos: ptoda} given in \ref{ldosptoda}. The plots of $\varrho(\lambda)$ looks Gaussian. For comparison, we plot the Gaussian distribution (black dashed line) with the variance equal to $T$ and to quantify the deviation from Gaussianity of the actual density we also quote the value of the Binder cumulant $U_4 = 1-\frac{\langle \lambda^4 \rangle}{3\langle \lambda^2 \rangle}$ for the Lax DOS of the cosh-confined Calogero fluid with $N=256$. }
\label{fig:todalowd}
\end{figure}

In Fig.~\ref{fig:todalowd}, we compare the Calogero Lax DOS  and the Toda chain with periodic boundary conditions, respectively confined by a hard box. The densities are $\bar{\rho} = 0.05, 0.075, 0.1$, all of which can be considered as relatively low density.  The temperature is $T=1$ and system sizes are $N= 128, 256$. In all cases, we find good agreement between the two models. This agreement becomes better for a lower particle density $\bar{\rho}$. As a comparison, we also display the exact infinite volume DOS obtained from Eq.~\eqref{dos: ptoda} given in~\ref{ldosptoda}.

\noindent
\textit{High density:} As explained above, in this limit 
the Calogero Hamiltonian can be  approximated by 
\begin{align}\label{rcm: E}
H_{\mathrm{r}}\left(\{q_i, p_i\}\right) = \sum_{j=1}^N \tfrac{1}{2} p_j^2+  \sum_{1 \leq i < j}^N \frac{1}{(q_i-q_j)^2 },
\end{align}
which is known as the rational Calogero model \cite{calogero1969solution, calogero1969ground, calogero1971solution}. This model is also integrable with the Lax pair
\begin{align}\label{Lax: rcm} 
\begin{split}
[L_{\mathrm{r}}]_{ij} &= \delta_{ij}p_j + \mathrm{i}(1- \delta_{ij}) \frac{1}{q_i- q_j},\\
[M_{\mathrm{r}}]_{ij} &=  \mathrm{i}\delta_{ij} \sum_{k = 1,k\neq j}^N\frac{1}{(q_j- q_k)^2} - \mathrm{i}(1- \delta_{ij})
\frac{1}{(q_i- q_j)^2}.
\end{split}
\end{align}
In Ref.~\cite{bogomolny2009random}, the spacing distribution of the eigenvalues of $L_\mathrm{r}$ was investigated for a special choice of the equilibrium Boltzmann weight where the energy $\mathrm{tr}(L_\mathrm{r}^2)$ in the Boltzmann weight is replaced by $\mathrm{tr}(L_\mathrm{r}^2) + \sum_{i=1}^N q_i^2$. 
Here, we consider the thermal Lax DOS when particles are confined by the cosh potential. It is of particular interest to compare the Lax DOS  with periodic boundary conditions. The periodized interaction potential can be obtained by using the following identity
\begin{align}
    \left(\frac{\pi}{\ell}\right)^2 \frac{1}{\sin^2\left(\tfrac{\pi}{l}r\right)} = \sum_{n=-\infty}^{\infty}\frac{1}{(r+n \ell)^2}.
\end{align}
Hence the periodized Hamiltonian is given by 
\begin{align}
    H_{\mathrm{t}}\left(\{q_i, p_i\}\right) = \sum_{i=1}^N \frac{p_i^2}{2} + \sum_{j>i\geq 1}^N \left(\frac{\pi}{\ell}\right)^2 \frac{1}{\sin^2\left(\tfrac{\pi}{l} (q_i-q_j)\right)}.
\end{align}
The trigonometric Calogero fluid~\cite{sutherland1971exact} is also integrable and possesses the Lax matrix structure~\cite{moser1975} with the following Lax pair
\begin{align}
    [L_{\mathrm{t}}]_{ij} &= \delta_{ij} p_i + {\rm i} (1-\delta_{ij})\left(\frac{\pi}{\ell}\right)\frac{1}{\sin\left(\tfrac{\pi}{\ell}(q_i-q_j)\right)},\\
    [M_{\mathrm{t}}]_{ij} &= {\rm i}\delta_{ij}\sum_{k =1, k\neq i}^{N}\left(\frac{\pi}{\ell}\right)^2\frac{1}{\sin^2\left(\tfrac{\pi}{\ell}(q_i-q_j)\right)} - {\rm i}(1-\delta_{ij})\left(\frac{\pi}{\ell}\right)^2\frac{\cos\left(\frac{\pi}{\ell}r\right)}{\sin^2\left(\tfrac{\pi}{\ell}r\right)}.
\end{align}
The Lax DOS for the trigonometric Calogero model has been obtained in Ref.~\cite{choquardclassical, spohn2023hydrodynamic} and we briefly discuss this result in \ref{trig}.
\begin{figure}[htb!]
\centering
\includegraphics[scale=0.7]{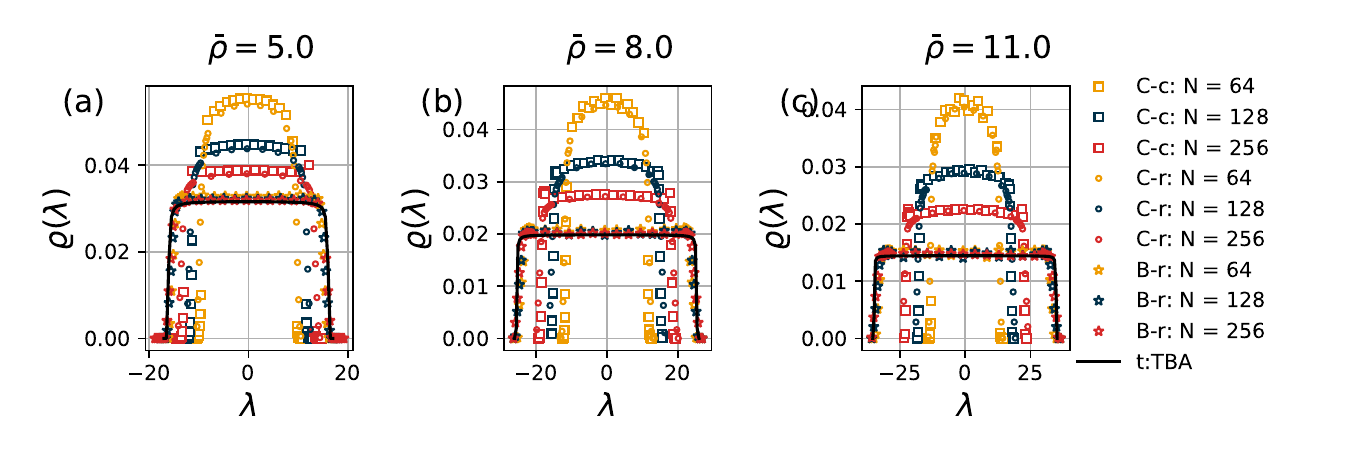}
\caption{The plot shows the Lax matrix DOS of the cosh-confined Calogero model (square symbol) and compares it to the rational Calogero model with either cosh potential (circle symbol) or box potential (star symbol). In addition, we display the exact DOS of the trigonometric Calogero model based on the TBA equation, see~\ref{trig}. The values of the fairly high densities are  (a) $\bar{\rho} = 5.0$, (b) $\bar{\rho} = 8.0$ and (c) $\bar{\rho} = 11.0$ for $T=1$ and with $N=64, 128, 256$. The average is over $10^5$ samples. The symbols ``C'' and ``B'' represent the cosh and box confinement respectively. The lowercase symbols ``c'', ``r'' and ``t'' represent the Calogero, the rational Calogero and the trigonometric Calogero fluid, respectively. }
\label{fig:comparratsinh}
\end{figure}

In Fig.~\ref{fig:comparratsinh}, we compare the Lax DOS of the cosh-confined Calogero fluid with the rational Calogero model subject to different boundary conditions: (i) external cosh confinement, (ii) box confined and (iii) on a ring. As discussed above the rational Calogero model on a ring corresponds to the trigonometric Calogero model where the analytical solution for the Lax DOS is well studied and we only plot the analytical solution for the case of the ring. The plot shows the Lax DOS of the corresponding models for three values of $\bar{\rho} = 5.0, 8.0, 11.0$, all of which are considered as relatively high density. We set the temperature to $T=1$ and consider system sizes $N=128 $ and $ 256$. The Lax DOS of the cosh-confined Calogero and rational Calogero models shows a very good agreement. However, neither of them have reached their thermodynamic limit. In this figure, we further demonstrate that with increasing $N$ the Lax DOS tend to be that of the TBA solution for the trigonometric Calogero fluid. On the other hand, the Lax DOS of the rational Calogero model in a box potential and periodic ring shows a very good agreement and the thermodynamic limit is also already achieved. 
\begin{figure}[b!]
    \centering
    \includegraphics[width=0.9\linewidth]{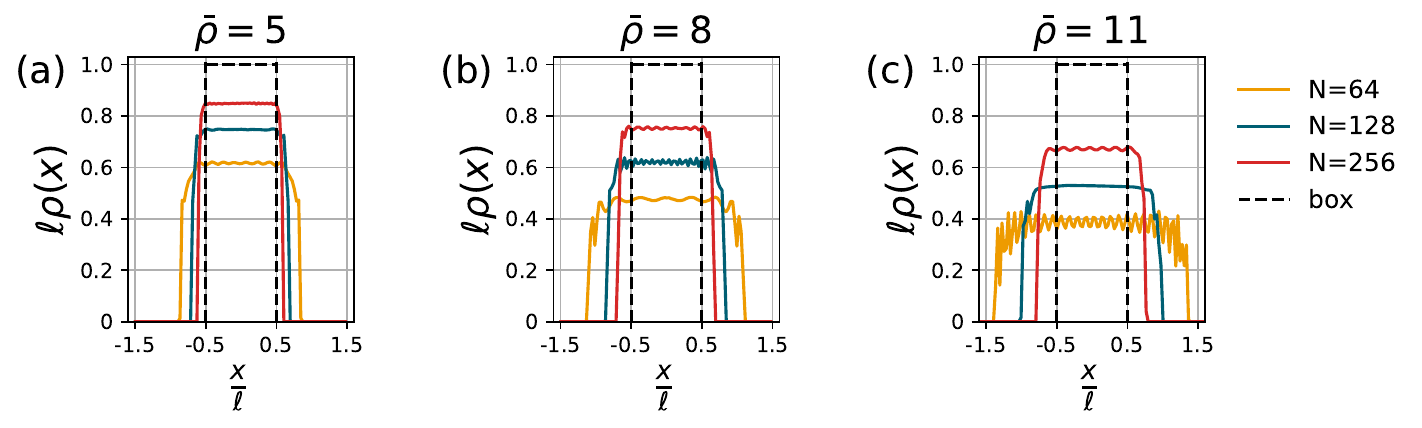}
    \caption{The plot shows the scaled particle density profile $\ell \rho(x)$ as a function of $x/\ell$ for cosh-confined Calogero fluid at $\bar{\rho} = 5, 8$ and $11$ with $N = 64, 128$ and $256$. Increasing $\bar{\rho}$, we observe slow convergence to a flat density profile (dashed line), which is expected for box-confined Calogero fluid.}
    \label{fig:rdos}
\end{figure}
There seem to be two effects whose importance is difficult to disentangle. 
Presumably, at high density the bulk DOS converges more slowly to its infinite size limit. On the other hand, since we compare with the rational model only, the slow convergence of the cosh-confined Calogero fluid and rational Calogero model could be merely a finite-size effect. To achieve a comparable convergence for the  DOS at $\bar{\rho} = 5.0$ in Fig.~\ref{fig:comparratsinh}a and for $\bar{\rho} = 11.0$ in Fig.~\ref{fig:comparratsinh}c one should compare using the same $\ell$. But for the denser system, this would require to study much larger system sizes which is computationally expensive. A more direct way of observing the discrepancy between box potential and cosh-confined potential is provided in Fig.~\ref{fig:rdos}. In this figure, we plot the density of cosh-confined Calogero particles and observe that the density profile of particles has not yet converged to flat density with support $[-\ell/2, \ell/2]$, as expected for the box confined Calogero fluid. 

\section{Conclusions and Outlook}
\label{sec:conc}

To summarize, we study a new class of random matrices which arise from the Lax matrix structure of classical integrable systems.  As in  Random Matrix Theory (RMT), the density of states (DOS), i.e. the density of eigenvalues, is the first and most natural quantity that is investigated. For the Calogero model confined in a box, we thoroughly investigate the DOS of the random matrix that emerges from its Lax matrix constructed from the phase space configurations sampled from the Boltzmann-Gibbs distribution. We demonstrate how the shape depends on the density of Calogero particles and their temperature, see Fig.~\ref{fig:dos_vs_T_rbar}. Remarkably, as demonstrated in Ref.~\cite{spohn2023hydrodynamic}, if the particles of the Calogero fluid are confined in an external cosh potential, the joint probability distribution of the eigenvalues of the Lax matrix of the Calogero fluid [Eq.~\eqref{Lax: hcm}] can be mapped to a modification of the conventional log-gas at high temperatures  [Eq.~\eqref{poflambda0}]. In this work, we have provided compelling numerical verification of this via extensive MC simulations. In Fig.~\ref{fig:comparbessinh}, we show an excellent match between the density of eigenvalues  obtained from  the modified log-gas and from directly diagonalising the Lax matrix of the Calogero fluid. Next, we address the question about the sensitivity of the Lax DOS to the details of the boundary conditions. 
To do so we consider three different boundary conditions: (i) an external trap of cosh type [Eq.~\eqref{vex}], (ii) the confining box with hard walls and (iii) on a ring. We provide compelling numerical evidence that in the thermodynamic limit, the DOS is independent of the specific boundary conditions used, see Fig.~\ref{fig:comparsinhweier}. This finding is consistent with the intuition that for the short-range models such as the Calogero fluid the role of boundary conditions is unimportant due to fast decaying correlations. 

The Hamiltonian for the Calogero fluid at low and high density limits corresponds to that of well-studied integrable models namely the Toda chain and the rational Calogero model, respectively. Consequently, one would expect the Lax DOS to match the respective limits, thus providing a nice procedure to extract the Lax DOS of these various deeply connected integrable models. In Fig.~\ref{fig:todalowd}, we present a comparison of the Lax DOS of the cosh-confined Calogero fluid in the low density of particles limit with that of the Toda model where we observe a good match. Similarly, in Fig.~\ref{fig:comparratsinh} we compare the Lax DOS of the cosh-confined Calogero fluid at a relatively high density with that of the cosh-confined rational Calogero model where we again observe a good match. In both cases, we study various boundary conditions and establish insensitivity to the details of the boundary condition in the thermodynamic limit. 

Our work is an important step forward in two main ways. Firstly, it puts forward a new class of random matrices arising from integrable models which may be amenable to analytical results for a variety of conventional random matrix diagnostics. For example, one could further explore quantities such as level spacing statistics~\cite{forrestor2010, haake1991, mehta2004}, adjacent gap ratio~\cite{huse2007} and spectral form factor~\cite{prakash2021,prasad2023, cotler2017, gharibyan2018, liu2018} to unravel the signatures of short and long-range correlations between eigenvalues in this new class of random Lax matrices. Secondly, the random Lax matrices, we investigated are also an important building block for generalized hydrodynamic descriptions of integrable systems. Although, in this work, we focused on an integrable particle model, it would be interesting to explore integrable lattice discretizations of continuum integrable field theories that possess Lax pair structure. Examples of such promising platforms are the Ablowitz-Ladik equation~\cite{ALbook,spohn2022} and integrable discretization of the Landau-Lifshitz model \cite{LL1,LL2,LL3}.

\section{Acknowledgements}

M. K. thanks the hospitality of the Department of Mathematics of the Technical University of Munich, Garching (Germany). M. K. and H. S. thank the VAJRA faculty scheme (No. VJR/2019/000079) from the Science and Engineering Research Board (SERB), Department of Science and Technology, Government of India. A. K. would like to acknowledge the support of DST, Government of India Grant under Project No. CRG/2021/002455 and the MATRICS grant MTR/2021/000350 from the SERB, DST, Government of India. J. K., M. K.  and A. K. acknowledge the Department of Atomic Energy, Government of India, for their support under Project No. RTI4001.
\appendix

\begin{center}
\line(1,0){250}
\end{center}

\section{Lax pair for the elliptic Calogero model}
\label{applaxwiers}
In this appendix, we summarize the construction of the Lax Matrices $L$ and $M$ for the elliptic Calogero model \cite{calogero2014}. The Lax matrices $L$ and $M$ have a general form
\begin{align}
 [L]_{ij} &= \delta_{ij}p_j + (1- \delta_{ij}) \alpha(q_i-q_j) ,\\
[M]_{ij}& = \delta_{ij} \sum_{k = 1,k\neq j}^N \eta(q_i-q_k)   + (1- \delta_{ij})\gamma(q_i-q_j).
\end{align}
The matrix evolution equation 
\begin{align}
   \dot{L} = [L, M]
   \label{app:ldot}
\end{align}
is equivalent to the Hamilton's equation of motion 
\begin{align}\label{eom}
\dot{q}_i = p_i,~~~\text{and}~~
\dot{p}_i = -\sum_{j \neq i}^N \tfrac{\partial}{\partial q_i}V(q_i-q_j),\,
\end{align}
where $V(r)$ is the interaction potential. Taking the time derivative of the $L$ matrix we find
\begin{align}
        \dot{[L]}_{ij} &= \delta_{ij}\dot{p}_j + (1- \delta_{ij}) \alpha'(q_i-q_j)(\dot{q}_i-\dot{q}_j),
\end{align}
where $\alpha'(r) = \tfrac{d}{dr}\alpha(r)$. The right-hand side of the Lax equation of motion given in Eq.~\eqref{app:ldot} is given by
\begin{align}
[LM-ML]_{ij}& = \delta_{ij}\left( \sum_{k \neq i} \left[\alpha(q_i-q_k)\gamma(q_k-q_i)-\alpha(q_k-q_i)\gamma(q_i-q_k)\right]\right)\notag\\
    &+(1-\delta_{ij})\alpha(q_i-q_j)\left[\eta(q_j-q_i)-\eta(q_i-q_j)\right]\notag\\
    &+(1-\delta_{ij})\left(p_i-p_j\right)\gamma(q_i-q_j)\notag\\
    &+(1-\delta_{ij})\Bigg(\sum_{k \neq i \& j}\Big\{\alpha(q_i-q_j)\left[\eta(q_j-q_k)-\eta(q_i-q_k)\right]\notag\\
    &+\alpha(q_i-q_k)\gamma(q_k-q_j)-\gamma(q_i-q_k)\alpha(q_k-q_j)\Big\}\Bigg).
\end{align}
Now using Hamilton's equation of motion given in Eq.~\eqref{eom} on the left-hand side of Eq.~\eqref{app:ldot} and comparing the off-diagonal terms we obtain
\begin{align}\label{app:gamma(r)}
 \gamma(q_i-q_j) &= \alpha'(q_i-q_j),\\
 \eta(q_i-q_j) &=\eta(q_j-q_i),\\
\eta(q_i-q_k)-\eta(q_k-q_j)&= \frac{\alpha(q_i-q_k)\alpha'(q_k-q_j)-\alpha'(q_i-q_k)\alpha(q_k-q_j)}{\alpha(q_i-q_j)}.
\end{align}
While comparing the diagonal entries of the left and right sides of Eq.~\eqref{app:ldot} and using Eq.~\eqref{app:gamma(r)} gives
\begin{align}
    -\frac{\partial}{\partial q_i}V(q_i-q_r) &= -\frac{\partial}{\partial q_i}\alpha(q_i-q_r)\alpha(q_r-q_i),\\
    V(q_i-q_r)+C_1 &= \alpha(q_i-q_r)\alpha(q_r-q_i).
\end{align}
where $C_1$ is a specific constant determined by $V(r)$ and $\alpha(r)$.
The relation between the functions $\alpha(r), \beta(r)$ and $\gamma(r)$ are summarized as follows:
\begin{align}\label{alpha(r)}
    \gamma(r) &= \alpha'(r),\\
    \eta(r) &= \eta(-r)\\
    \label{app:eta(r)} \eta(r)-\eta(u) & = \frac{\alpha(r)\alpha'(u) - \alpha(u)\alpha'(r)}{\alpha(r+u)}\\
   V(r) + C_1 &= \alpha(r)\alpha(-r).
\end{align}
Analyzing the Eq.~\eqref{app:eta(r)} when $r + u =\epsilon \ll 1$ by comparing terms order by order as a series in $\epsilon$, we find that $\alpha(-r) = -\alpha(r)$ and \begin{align}\label{eta(r)-cl}
\eta(r) = -\frac{V(r)}{b_{-1}}+c_2    
\end{align}
where $c_2$ is any arbitrary constant and $b_{-1}$ is obtained from \cite{calogero2014}
\begin{align}
    \alpha(r\to 0) = \frac{b_{-1}}{r} + b_0 +b_1 r +b_3 r^3 +O(r^5).
\end{align}
The arbitrary nature of $c_2$ is because $M \to M +I c_2$ does not modify the equation of motion [Eq.~\eqref{app:ldot}] with $I$ being the identity matrix with the same dimension as $M$.  

For the periodized Calogero fluid [see Eq.~\eqref{weier: PE}]  the functions $\alpha_{\mathrm{w}}(r) \equiv \alpha(r)$ and $\beta_{\mathrm{w}}(r) \equiv \beta(r)$ can be obtained by noting that the periodic Weierstrass function and the Jacobi Sin and Jacobi Cos function are related by the identity \cite{korn1968, calogero2014}
\begin{align}\label{identity2}
    \wp(r|w,w')-e_1 = \left(\frac{\sqrt{e_1-e_3}~\mathrm{Cn}(r\sqrt{e_1-e_3}|m)}{\mathrm{Sn}(r\sqrt{e_1-e_3}|m)}\right)^2.
\end{align}
Here the Jacobi Sin and Jacobi Cos functions are defined as 
\begin{align}\label{app:jacobi}
    \mathrm{Sn}(r,m) = \sin(\phi(r,m)),~~\text{and}~~   \mathrm{Cn}(r,m) = \cos(\phi(r,m)).
\end{align}
where the Jacobi amplitude $\phi(r,m)$ satisfies 
\begin{align}
    r = F(\phi, m) = \int_{0}^{\phi}d\theta \frac{1}{\sqrt{1-m^2\sin^2 \theta}}.
\end{align}
For notational simplicity in Eq.~\eqref{identity2} and rest of the article we use $e_i \equiv e_i(w, w')$ with $i=1,2,3$ and $m \equiv m(w, w')$
\begin{align}
    m(w, w') &= \sqrt{\frac{e_2-e_3}{e_1-e_3}}~~\text{where}~~e_1(w, w') = \wp(w|w, w'),\\
    e_2(w, w') &= \wp(-w-w'|w, w')~~\text{and}~~e_3(w, w') = \wp(w'|w, w')\label{ei}.
\end{align}
Now, setting $V(r)+C_1 = V_{\mathrm{w}}(r)$ in Eq.~\eqref{alpha(r)} and using the identity in Eq.~\eqref{identity2} we find that
\begin{align}\label{alpha_w}
    [\alpha_{\mathrm{w}}(r)]^2 &= - V_{\mathrm{w}}(r)=-\left(\wp(r|w,w')-e_1\right)= - \left(\frac{\sqrt{e_1-e_3}~\mathrm{Cn}(r\sqrt{e_1-e_3}|m)}{\mathrm{Sn}(r\sqrt{e_1-e_3}|m)}\right)^2,
\end{align}
where we have used $\alpha_{\mathrm{w}}(-r) = -\alpha_{\mathrm{w}}(r)$. Recall that the subscript ``$\mathrm{w}$'' in Eq.~\eqref{alpha_w} represents the elliptic Weierstrass model. Using Eq.~\eqref{alpha_w}, we find that
\begin{align}
\alpha_{\mathrm{w}}(r) = \mathrm{i}\frac{\sqrt{e_1-e_3}~\mathrm{Cn}(r\sqrt{e_1-e_3}|m)}{\mathrm{Sn}(r\sqrt{e_1-e_3}|m)}.
\end{align}
The function $\eta_{\mathrm{w}}(r)$ is obtained by using Eq.~\eqref{eta(r)-cl} where the coefficient $b_{-1}$ is obtained by using the fact that
\begin{align}
    \alpha_{\mathrm{w}}(r \to 0) = \frac{\mathrm{i}}{r} +O(r),
\end{align}
which gives $b_{-1} = \mathrm{i}$.

In our numerical simulations, we set 
\begin{align}\label{wwp}
w = \frac{\ell}{2}~~\text{and}~~ w' = \mathrm{i} \pi 
\end{align}
and use the Eq.~\eqref{identity2} to relate the function $\alpha_{\mathrm{w}}(r)$ in terms of the periodic Weierstrass function as
\begin{align}\label{alphaw(r)}
    \alpha_{\mathrm{w}}(r) =\begin{cases}
\mathrm{i}~\sqrt{\wp\left(r\bigg|\frac{\ell}{2}, \mathrm{i}\pi\right)-    \wp\left(\frac{\ell}{2}\bigg|\frac{\ell}{2}, \mathrm{i} \pi\right)}~&\text{for}~r>0\\
-~\mathrm{i}~\sqrt{\wp\left(r\bigg|\frac{\ell}{2}, \mathrm{i} \pi\right)-    \wp\left(\frac{\ell}{2}\bigg|\frac{\ell}{2}, \mathrm{i} \pi\right)}~&\text{for}~r<0
    \end{cases},
\end{align}
where the function $\wp\left(r\bigg|\frac{\ell}{2}, \mathrm{i}\pi\right)-    \wp\left(\frac{\ell}{2}\bigg|\frac{\ell}{2}, \mathrm{i} \pi\right)$ is approximated by truncating Eq.~\eqref{sinhtowp} at $n = \pm 2$ {\it i.e.}
\begin{align}\label{app:sinhtowp}
    \wp\bigg(r \bigg|\frac{\ell}{2}, \mathrm{i} \pi \bigg)-{\wp\bigg(\frac{\ell}{2}\bigg|\frac{\ell}{2}, \mathrm{i} \pi \bigg)} = \sum_{n=-2}^{2}\left[\frac{1}{4\sinh^{2}\left(\frac{r}{2} + n\ell\right)}-{\frac{1}{4\sinh^{2}\left(\frac{\ell}{4} + n\ell\right)}}\right].
\end{align}

\section{Lax DOS of the Toda Chain}
\label{ldosptoda}
In this appendix, we discuss the Lax DOS of the Toda chain when the system is in thermal equilibrium at inverse temperature $\beta$ and constant pressure $P$. Such ensemble was studied in Ref.~\cite{spohn2020generalized}, where it was shown that the Lax DOS is related to the following free energy functional 
\begin{align}\label{free: ptoda}
    \mathcal{F}_{P}(\rho) = \int_{-\infty}^{\infty}&d\lambda~\tfrac{1}{2}\beta\lambda^2\rho(\lambda)+ \log P-P \int_{-\infty}^{\infty}\int_{-\infty}^{\infty}d\lambda~d\lambda'~\log\left(|\lambda-\lambda'|\right)\rho(\lambda)\rho(\lambda') \notag\\&+ \int_{-\infty}^{\infty}d\lambda~\rho(\lambda)\log\rho(\lambda).
\end{align}
This is the free energy of the conventional log-gas. In RMT one uses a scaling
such that energy dominates entropy. However, in Eq.~\eqref{free: ptoda} energy and entropy are of the same order. In the terminology of RMT, Eq.~\eqref{free: ptoda} then corresponds to high temperatures.

The free energy Eq.~\eqref{free: ptoda} is derived for the Toda chain with a linear pressure ramp. Thus to obtain the Toda DOS one first has to minimize Eq.~\eqref{free: ptoda} over all normalized probability densities $\rho$. Denoting this unique minimizer by $\rho^*(\lambda)$, 
the Toda Lax DOS, $\varrho(\lambda)$, is given by
\begin{align}\label{dos: ptoda}
    \varrho(\lambda) = \partial_P\left(P\rho^*(\lambda)\right).
\end{align}
In fact, in the case of thermal equilibrium, an exact expression for $\rho^*$ has been derived in ~\cite{allez2012, opper1985} with the result
\begin{align}\label{ptoda: theory}
    \rho^*(\lambda) = \frac{1}{\sqrt{2 \pi} \Gamma(1+P)}\frac{1}{|D_{-P}(\mathrm{i}\lambda)|^2}
\end{align}
where the function $D_{-P}(z)$ has the following integral representation
\begin{align}
    D_{-P}(z) = \frac{1}{\Gamma(P)}\exp\left(-\tfrac{1}{4}z^2\right)\int_{0}^{\infty}dx~x^{P-1}\exp\left(-z\,x-\tfrac{1}{2}x^2\right).
\end{align}

To compare the Lax DOS of the Toda chain [Eq.~\eqref{dos: ptoda}] under constant pressure and the confined Toda chains with fixed particle density  $\bar{\rho} = \frac{1}{\nu}$ we use the following relation to compute the pressure 
\begin{align}\label{relationpressurestretch}
\frac{1}{\bar{\rho}} = \nu = \langle r\rangle_{P} = \frac{\partial \mathcal{F}_{{\mathrm{to}}}}{\partial P},\qquad \mathcal{F}_{{\mathrm{to}}} = \partial_{P}\left(P\mathcal{F}_{P}(\rho^*)\right)-(1+2P)\log(2).
\end{align}
Recall that the subscript ``$\mathrm{to}$'' in Eq.~\eqref{relationpressurestretch} represents the Toda interaction.

\section{Lax DOS of the trigonometric Calogero model}
\label{trig}
We discuss the Lax DOS of the trigonometric Calogero model when the system is in thermal equilibrium at inverse temperature $\beta$ and confined to a ring of length $\ell$ through periodic boundary conditions. This model was studied in Refs.~\cite{choquardclassical,spohn2023hydrodynamic}.
By using action-angle coordinates it is shown that the thermal Gibbs distribution can be expressed as  
\begin{align}\label{p-trig}
     \frac{1}{Z(\beta, \ell)}\exp\left(-\beta H_{\mathrm{t}}\left(\{q_i, p_i\}\right)\right)\prod_{i=1}^Ndp_idq_i
    =\frac{1}{Z(\beta, \ell)}\exp\left(-\tfrac{\beta}{2}\mathrm{tr}[L_{\mathrm{t}}^2]\right)\chi(\{\lambda_i\})\prod_{i=1}^Nd\lambda_id\phi_i,
\end{align}
where subscript ``$\mathrm{t}$'' represents the trigonometric Calogero model. Here $\{\phi_i\}$ are the angle variables taking values on the ring, $0<\phi_i<\ell$.  
The action variables, $\{\lambda_i\}$, are the eigenvalues of the Lax matrix $L_{\mathrm{t}}$. They are constrained such that $\lambda_{i+1}-\lambda_i\geq 2\pi/\ell$ for $1\leq i\leq N-1$, which can be expressed as 
\begin{align}
    \chi(\{\lambda_i\}) = \prod_{i=1}^{N -1}\Theta\left(\lambda_{i+1}-\lambda_i-\tfrac{2\pi}{\ell}\right),
\end{align}
using the Heavyside function $\Theta(x) = 1$ for $x>0$ and $\Theta(x)=0$ otherwise. The partition function can then be expressed as
\begin{align}
    Z(\beta, \ell) = \ell^N \int~\prod_{i=1}^Nd\lambda_i\exp\left(-\tfrac{\beta}{2}\lambda_i^2\right) \prod_{i=1}^N\Theta\left(\lambda_{i+1}-\lambda_i-\tfrac{2\pi}{\ell}\right),
\end{align}
which is exactly the thermal partition function at temperature $\beta^{-1}$ of hard rods of length $2\pi/\ell$ confined by a harmonic trap with trap strength $1$. As shown in \cite{spohn2023hydrodynamic}, the thermodynamic limit of the Lax DOS is given by
\begin{align}
\varrho^*(\lambda) = \frac{f^*(\lambda)}{(1+2\pi f^*(\lambda))},
\end{align}
where the function $f^*(\lambda)$ satisfies the following transcendental equation ~\cite{choquardclassical}
\begin{align}\label{saddle-trig}
    f^*(\lambda) = \exp\left(-2 \pi f^*(\lambda)\right)\exp\left(-\tfrac{\beta}{2}\lambda^2 + \mu\right).
\end{align}
The function  $f^*(\lambda)$ is the derivative of the free energy density of the harmonically trapped hard rods in the thermodynamic limit. Here the chemical potential $\mu$ is fixed through the normalization constraint, $\int d\lambda~\varrho^*(\lambda)=1$. In Fig.~\ref{fig:comparratsinh}, we compare the Lax DOS obtained by solving Eq.~\eqref{saddle-trig} with those obtained from MC simulations of the Calogero model and the rational Calogero model at high densities.

\end{document}